\documentclass[aps,pre,floatfix,onecolumn,showkeys,]{revtex4}
\usepackage{color}
\usepackage{float}
\usepackage{xcolor}
\usepackage{tabularx}
\usepackage{natbib}

\usepackage[colorlinks=true,urlcolor=magenta,citecolor=blue,linkcolor=blue]{hyperref}

\usepackage{graphicx} 
\usepackage{amsmath} 
\usepackage{amssymb}

 \definecolor{Gainsboro}{rgb}{0.86,0.86,0.86}
 
 \newcommand{\thickhline}{%
    \noalign {\ifnum 0=`}\fi \hrule height 1pt
    \futurelet \reserved@a \@xhline
}
\newcolumntype{"}{@{\hskip\tabcolsep\vrule width 1pt\hskip\tabcolsep}}

\newcommand{\RNum}[1]{\uppercase\expandafter{\romannumeral #1\relax}}

\begin{document}

\title{Statistical tests for MIXMAX pseudorandom number generator}
\author{Narek H. Martirosyan}
\email{narek.h.martirosyan@gmail.com, narek@mail.yerphi.am}
\affiliation{Yerevan Physics Institute, 2 Alikhanian Brothers street, Yerevan
0036, Armenia}
\author{Gevorg A. Karyan}
\affiliation{Yerevan Physics Institute, 2 Alikhanian Brothers street, Yerevan
0036, Armenia}
\author{Norayr Z. Akopov}
\affiliation{Yerevan Physics Institute, 2 Alikhanian Brothers street, Yerevan
0036, Armenia}

\begin{abstract}
  The Pseudo-Random Number Generators (PRNGs) are key tools in Monte Carlo simulations. More recently, the MIXMAX PRNG has been included in ROOT and Class Library for High Energy Physics (CLHEP) software packages and claims to be a state of art generator due to its long period, high performance and good statistical properties. In this paper the various statistical tests for MIXMAX are performed. 
The results compared with those obtained from other PRNGs, e.g. Mersenne Twister, Ranlux, LCG reveal better qualities for MIXMAX in generating random numbers. 
The Mersenne Twister is by far the most widely used PRNG in many software packages including packages in High Energy Physics (HEP), however the results show that MIXMAX is not inferior to Mersenne Twister.

\end{abstract}


\maketitle

\section{\textbf{ Introduction.}} 

In recent years, there is a growing interest on PRNGs in different branches of physics and not only.
A good PRNG is important to have guaranteed results of Monte Carlo (MC) methods. There are many software packages for MC simulations where PRNGs are the central components.  Among these packages one can mention the Geant4/CLHEP \cite{geant4}, a widely used simulation toolkit in HEP for modeling the passage of elementary particles through matter, also used for medical and space science simulations. 

PRNGs are also crucial in Markov Chain Monte Carlo (MCMC) methods  which are used for sampling from desired probability distribution by constructing Markov chain on state space whose stationary distribution is of interest \cite{Bharucha,MCMC1_Metropolis,MCMC2_Metropolis}. Uniform PRNGs play a central role in constructing such Markov Chains. 
Most of MCMC algorithms are developed within random walk models. 

A widely used example of random walk Monte Carlo method is  Metropolis-Hastings algorithm \cite{MCMC1_Metropolis,MCMC2_Metropolis,metropolis1, metropolis2} which is also included in the list of the top 10 algorithms \cite{Top10}. MCMC methods are mainly used for sampling from large dimensional spaces and computing multidimensional integrals. 
For example, in statistical mechanics, one needs to compute thermal averages of quantities, such as the total energy, magnetization, etc. by performing multidimensional integration or summation over configuration space. However, the total number of 
configurations can be very large, e.g. in 3-dimensional Ising model the number of spin configurations with particles at $n^3$ lattice sites is $2^{n^3}$. In thermodynamic equilibrium  the probabilities of occurring each configuration is represented by Boltzmann distribution. Thereby having samples drawn from Boltzmann distribution one can compute expectation values of thermodynamic quantities.

The necessity to have large amounts of simulated data imposes a strict requirements on PRNGs, such as statistical properties of generated numbers, swiftness in number generation, replicability, lengthiness of generated random cycle and independence of produced random numbers. 

To address these challenges the renewed version of MIXMAX PRNG \cite{sav1,sav2} based on Anosov C-systems and Kolmogorov K-systems has been introduced and developed in \cite{KS_2016,sav3, MIXMAX_2016,Kostya,sav_kostya}. The MIXMAX is matrix-recursive PRNG and it has been shown that the properties of the MIXMAX generator is improved with increasing the size $N$ of MIXMAX matrix \cite{Kostya}. The period of MIXMAX is also increased with increasing $N$ and it can be reach up to $10^{57824}$, note that the period of commonly used version of Mersenne Twister based on Mersenne prime has the period of $2^{19937}-1$.


While having a long period, however statistical properties and time characteristics of PRNGs are crucial to consider a generator "good" or "bad".
In this paper we will present the results of the statistical tests performed with the matrix size of $N=256$ which is considered to be a
default dimension of MIXMAX matrix with flexibility to be further increased. 

\section{\textbf{Visual demonstration.}} 

\begin{figure}[htbp]
\centering
    \includegraphics[width=9cm]{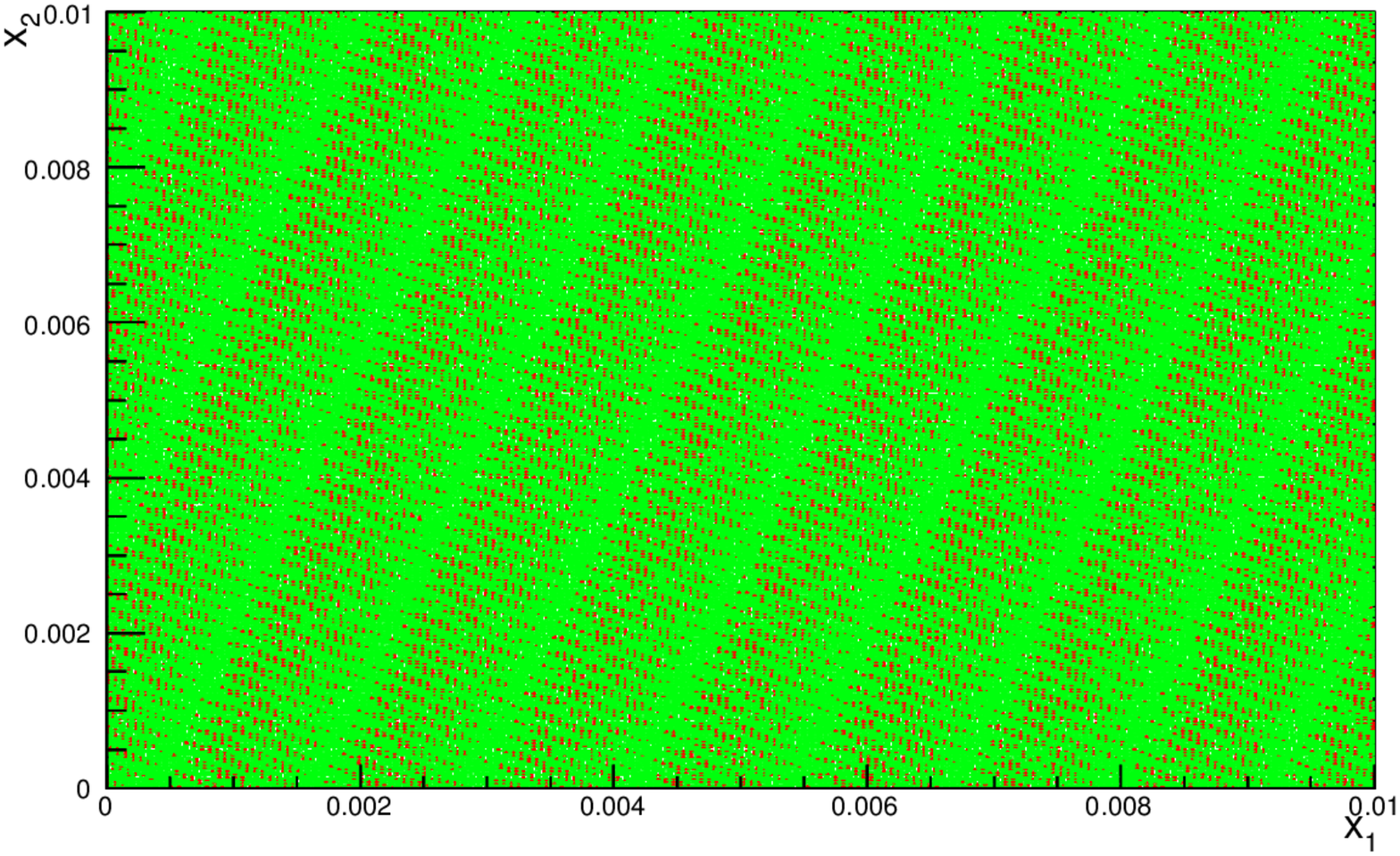}
    \includegraphics[width=9cm]{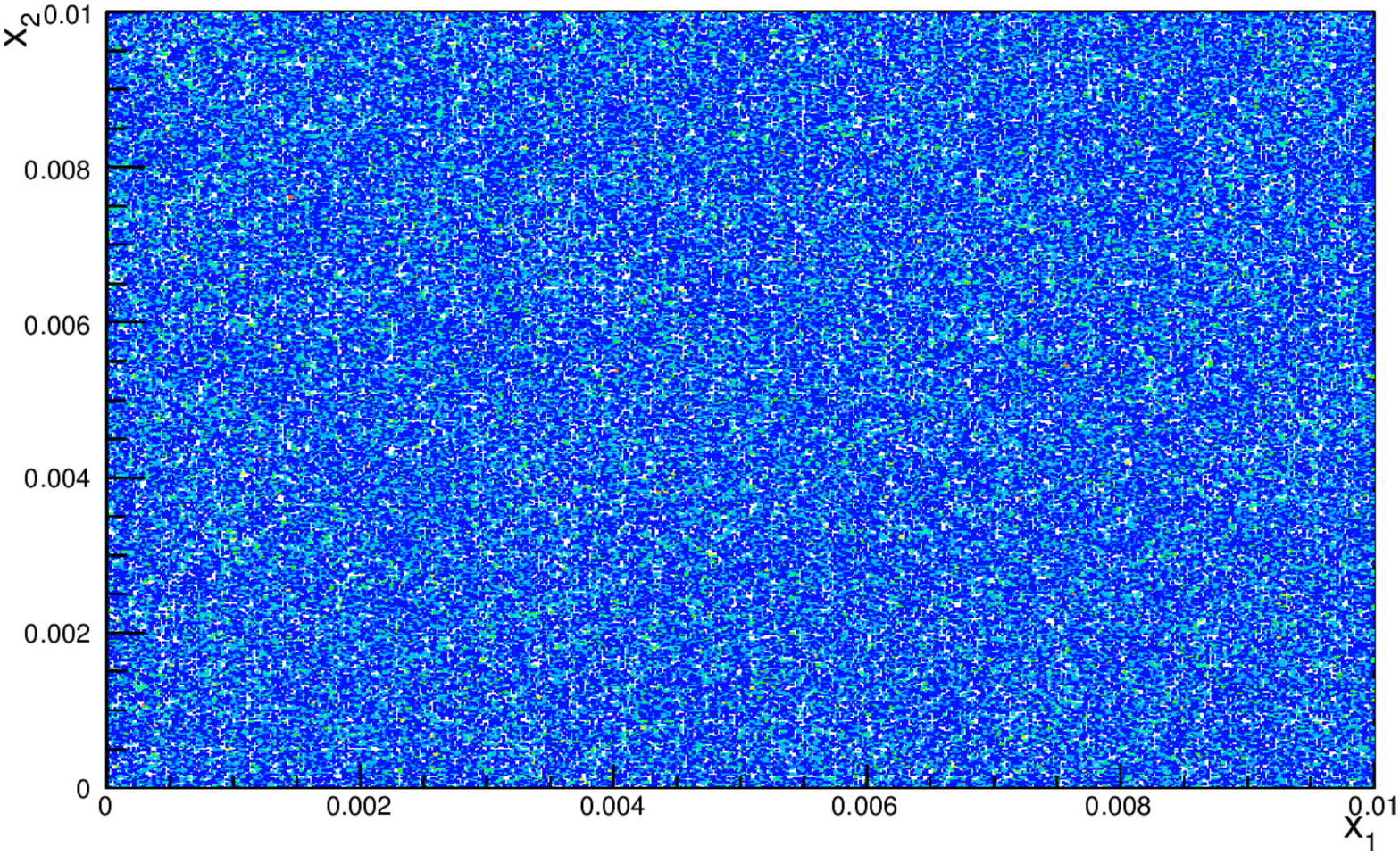}
    \caption{Random points in two-dimensional space generated by LCG (up) and MIXMAX (down)}
    \label{fig_visual}
\end{figure}

We can reveal the defect of uniform PRNGs simply plotting random points in high-dimensional Euclidean space, if these points form lattice structure then to a first approximation we can say that PRNG has defects in generating random points since the space is not filled uniformly. Fig.~\ref{fig_visual} shows the comparison of MIXMAX with the Linear Congruential Generator(LCG),  which is known to be defective PRNG. In contrast to LCG MIXMAX does not form lattice structure. We obtain these figures by generating two $U[0,1]$ random number sequences and assigning a point in two-dimensional space.

\section{\textbf{Statistical testing with TestU01.}} 

Most of PRNG algorithms produce numbers uniformly distributed in the interval of $(0,1)$, hence PRNGs should pass statistical tests of uniformity. Many empirical statistical testing packages implement tests for these purposes, some of which are \cite{Diehard,testu01,SPRNG, NIST}.
Most of the statistical tests implemented in these packages are discussed in Knuth's book \cite{Knuth}, e.g. the package \cite{SPRNG} implements mainly tests of Knuth. Currently, one of the well known tools for statistical testing is TestU01 software library which provides implementations of the empirical statistical tests for uniform PRNGs. It contains more than 160 different empirical tests and offers several batteries of tests including the most powerful one, i.e. the 'Big Crush'. When a specific statistical test is applied to random numbers produced by PRNG the {\it p}-value of the test is printed as a measure of deviation from null-hypothesis, which in our case is uniform distribution of random numbers. In comparison with other libraries TestU01 is more flexible and efficient, and it can deal with larger sample sizes and has wider range of test statistics than other libraries.

In the Table \ref{tab:1}, the outcome of TestU01 BigCrush suite applied on MIXMAX, Mersenne Twister and LCG is stored
by using 64-bit computer with $Intel \: Core \: i3$ $-4150$ processor of clock speed $3.50\times 4 \:GHz$.

As we can see from the table the MIXMAX passes the same test suite faster than Mersenne Twister and does not fail any test.
TestU01 test suite has been applied to Ranlux PRNG with its modifications Ranlux24, Ranlux48. It is  observed that Ranlux though having good statistical properties is very slow at generating random numbers.
Comparing with MIXMAX Ranlux24 is about 10 times slower and Ranlux48 is 17 times slower. This fact makes it not convenient for the use in generation of large amount of random numbers.   

\begin{table}[H]
\begin{center}
\vspace{0.3cm}
\begin{tabular}{|c"c |c| c|}
\hline
$\textcolor{gray}{}$ PRNG & Total CPU time & BigCrush& Failed test'(s) p value\\
\hline
$\textcolor{red}{\textbf{MIXMAX}}$ & $\textcolor{black}{2h \: 43m \: 51s}$ & $\textcolor{black}{All \:tests \:were \:passed}$ &
$\textcolor{black}{\line(1,0){10}}$\\
\hline
$\textcolor{blue}{\textbf{Mersenne Twister}}$ & $\textcolor{black}{3h \: 19m\: 27s}$ & $\textcolor{black}{3}$ &
$\textcolor{black}{0.9990,1-10^{-15}}$\\

\hline
$\textcolor{green}{\textbf{LCG}}$ & $\textcolor{black}{3h \: 30m \: 33s}$ & $\textcolor{black}{22}$ &
$\textcolor{black}{< 10^{-300}}$\\
\hline
\end{tabular}
\caption{TestU01 BigCrush suite results.} 
\label{tab:1}
\end{center}
\end{table}

\section{\textbf{Kolmogorov-Smirnov tests.}} 
Kolmogorov-Smirnov (K-S) test is one of the powerful tools that can be used to examine the statistical features of PRNGs. 

Though one-dimensional (1D) K-S test is already implemented in TestU01, we perform K-S test independently for various parameters of sample size ($n$) and extract the distribution of K-S test statistic.
The idea behind the test is to calculate maximum distance between
expected Cumulative Distribution Function(CDF) $F(x)$, $F(x) = Pr(X \leq x)$ and measured or Empirical Cumulative Distribution Function(ECDF) $F_{n}(x)$ of $n$ data points
\begin{equation}\label{ECDF}
F_{n}(x) = \frac{1}{n} (number~of~x_{i} \leq x)
\end{equation}

The null hypothesis $H_{0}$ is whether the sample of $n$ random numbers comes from expected distribution $F(x)$ or not. If data comes from $F(x)$, then the strong law of large numbers provides $F_{n}(x) \to F(x)$, as $n \to \infty$. The latter is strengthened by the Glivenko-Canteli (G-C) theorem \cite{Serfling}, which states that under $H_{0}$ hypothesis
\begin{equation}
Pr ( \lim_{n\to\infty} sup_{x} \: | F_{n}(x) - F(x) | = 0) = 1
\end{equation}

Hence the difference between CDFs can be used as a measure of agreement between a data and a given distribution. There are several statistical tests based on (G-C) theorem known as Cramer-von Mises tests \cite{Misses, Anderson}. The key feature of tests based on G-C theorem is that their distributions are independent of the hypothesized models under $H_{0}$ when data sample is large.

The one dimenisonal (1D) K-S test is defined as follows:
\begin{equation}\label{KsTestStatistic}
D_{n} = sup_{x} \: | F_{n}(x) - F(x) |
\end{equation}

Under null hypothesis the distribution of  $\sqrt{n}\cdot D_{n}$ converges to Kolmogorov distribution for sufficiently large $n$ when $F(x)$ is continuous \cite{Marsaglia}

\begin{equation}\label{KsDistr}
\lim_{n\to\infty}Pr(\sqrt{n}\cdot D_{n} \leq x) \equiv K(x) = 1 - 2\sum_{i=1}^{\infty}(-1)^{i-1} e^{-2i^{2}x^{2}}
\end{equation}
\noindent
It is of interest to note that for small values of $n$ Kolmogorov distribution is not adequate, but there is a way to compute  ${\it p}-value$ for randomly produced $D_{n}$ \cite{Marsaglia}.

\begin{figure}[H]
\centering
\includegraphics[width=10cm]{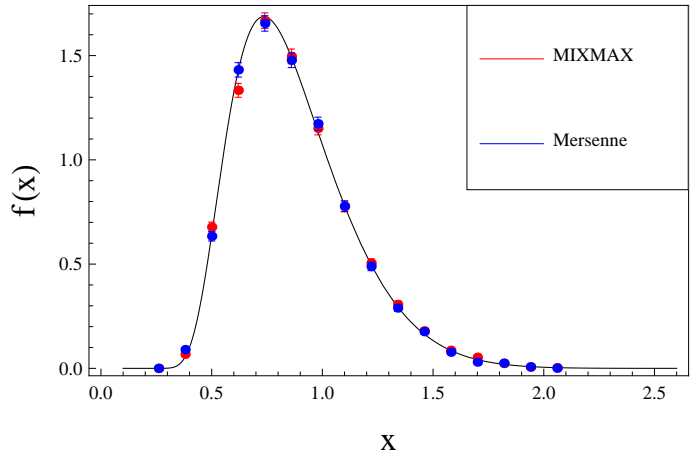} 
\caption {Distribution of $\sqrt{n}\cdot D_{n}$ for MIXMAX and Mersenne Twister. The size of a samples is $n = 10^8$ and the number of different replicas is $10^4$. The black curve is the PDF of Kolmogorov distribution.}
\label{fig3}
\end{figure}

In Fig.~\ref{fig3} the normalized histogram of $\sqrt{n}\cdot D_{n}$ data points for MIXMAX and Mersenne Twister is presented and compared with Probability Density Function (PDF) of Kolmogorov distribution: $f(x) = K'(x)$

\begin{equation}\label{PDFKolm}
f(x) = 8x \sum_{i=1}^{\infty}(-1)^{i-1} i^2 e^{-2i^{2}x^{2}}
\end{equation}

\begin{figure}[H]
\centering
\includegraphics[width=10cm]{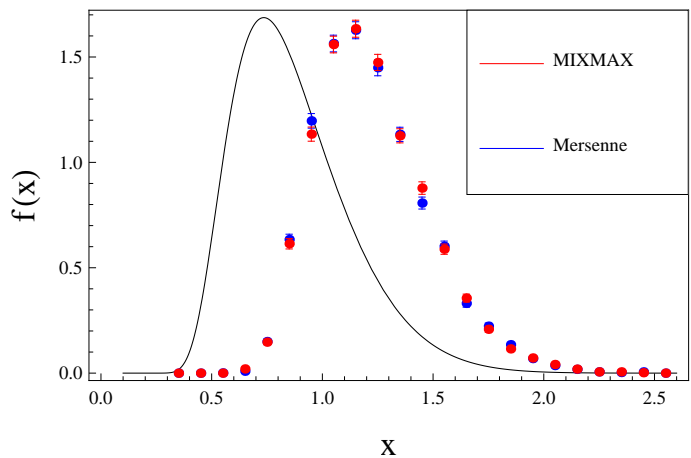} 
\caption {Distribution of $\sqrt{n}\cdot D_{n}$ for MIXMAX and Mersenne Twister for 2-dimensional case. The size of a samples is $n = 10^3$, note that $n$ is the total number of random points in 2-dimensional space, i.e. $10^6$ numbers are generated by PRNGs, the number of different replicas is $10^4$. The black curve is the PDF of Kolmogorov distribution. The shape of histogram shows clear shift from Kolmogorov distribution. MIXMAX and Mersenne Twister that have been used in these studies give almost the same distributions for $\sqrt{n}\cdot D_{n}$ which seems to be independent from dimension of Kolmogorov-Smirnov test.}
\label{fig4}
\end{figure}

\begin{figure}[htpb]
  \begin{center}
    \includegraphics[width=11.5cm]{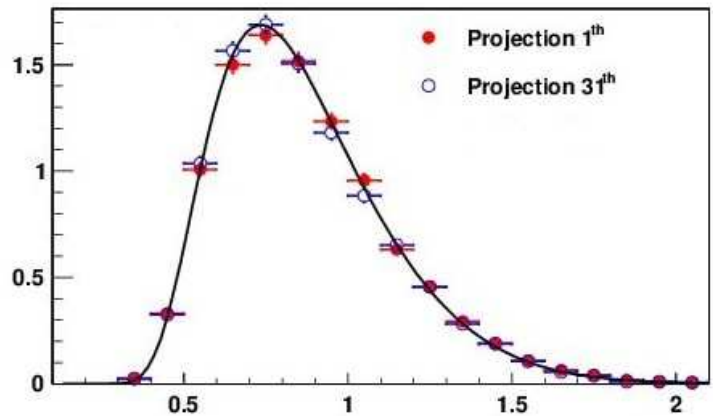}
    \caption{Distribution of $\sqrt{n}\cdot D_{n}$ for the $1^{th}$ and $31^{th}$ projections of MIXMAX in comparison with the theoretical expectation (black curve). The sample size is $n$ = $10^{8}$ and for each projection $10^{4}$ different replicas are generated.}
    \label{fig5}
  \end{center}
\end{figure}

The histograms in Figs.~\ref{fig3}--\ref{IrwinHallFigure} are normalized to unity dividing each bin entry by the product of sample size and bin width ($n \cdot width$). Visual comparison shows that under $H_{0}$ the distribution of $\sqrt{n}\cdot D_{n}$ follows the PDF of Kolmogorov distribution.
Due to fast convergence of the series of partial sums in Eqs.(\ref{KsDistr},\ref{PDFKolm}) it suffices to take limited number of terms in the sum, e.g. the first 100 terms are enough. 

Two-level tests can be used on K-S test to give evidence of visual coincidence on Fig.~\ref{fig3}. For this purpose chi-square test(next section) is applied. It has been checked that both distributions agree with theoretical expectation (black curve) in 95\% Confidence Level (CL).



In multidimensional K-S test one have $d$-dimensional data ($d \geq 2$) and to test $H_{0}$ it is needed to compare $d$-variate ECDF with the hypothetical $d$-variate CDF. The complication in multidimensional case is caused by the ambiguity in definition of the CDF since there are $2^{d} - 1$ independent ways of defining CDFs. There have been proposed different 

\noindent
ways to calculate the multidimensional K-S statistic \cite{2dks_1, 2dks_2}. In \cite{2dks_1} four quadrants around all combinations $(x_{i}, x_{j})$ of data points is considered and $D$ is taken as the maximum of 4 differences between CDFs over all quadrants. Therefore this idea makes the test statistic independent of ordering the data. The number of all pairs $(x_{i}, x_{j})$ for N points is equal to $N^{2}$, therefore to calculate $D_{n}$ one needs to compute differences between CDFs in $3N^{2}$ quadrants(the probability for fourth quadrant is found from normalization). This method suffers from the computing time when $N$ is large. In \cite{2dks_2}, it is proposed to consider only observed points rather than all combinations thereby reducing the computational time by computing the differences for $3N$ quadrants only. It is possible to compute $D_{n}$ with computationally higher efficiency introducing a binning technique applied to a continuous multidimensional data, i.e. discretizing the data space. The idea of  binning technique is discussed in \cite{Narek}.

In \cite{2dks_3} the algorithm for 2D K-S test is presented when only one CDF from all possible configurations is taken into account. As a result of it, the procedure used to compute $D$ evaluating the difference of CDFs is reduced to a small number of data points. 

In our studies we have extended the standard definition of one dimensional cumulative distribution to its two dimensional $``$analogue$"$ and computed K-S statistic using algorithm presented in \cite{2dks_3}~(see Fig.~\ref{fig4}).

Three-dimensional extension of K-S test is presented in \cite{3d_KStest}, where it is considered 8 CDFs, and using MC techniques the table of critical values are also presented in this paper.

To detect possible non-uniformities in the multidimensional random sequences of MIXMAX PRNG, an arbitrary selected projection has been checked via
Kolmogorov-Smirnov test and the results are compared with that of the first projection. In Fig.~\ref{fig5}, the probability density distributions of
Kolmogorov-Smirnov statistic are presented for the $1^{th}$ and the $31^{th}$ projections and the comparison is provided with the expected
distribution.

\section{\textbf{Chi-Square tests.}} 

\begin{figure}[H]
  \begin{center}
    \includegraphics[width=10cm]{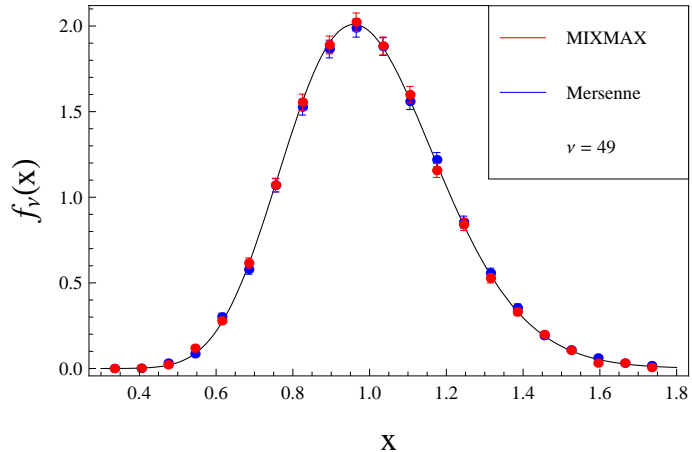}
\caption{Comparing denisty histogram of reduced chi-square for MIXMAX and Mersenne Twister with the PDF in (\ref{RedchisquareDist}). The size of a samples is $n = 10^6$ and the number of different replicas is $10^4$.}
  \label{fig6}
  \end{center}
\end{figure}

The chi-square $\chi^{2}$ test is one of the famous statistical tests which is met in many applications when one deals with grouped or binned data. The Chi-Square test is applied to categorical sample distributions unlike the K-S test which using each random point compares continuous sample distributions with hypothesized ones. The $\chi^{2}$ statistic has the following form \cite{Knuth,hudson}

\begin{equation}\label{chisquareStat}
\chi^{2} = \sum_{i=1}^{k}\frac{(O_{i} - E_{i})^2}{E_{i}},
\end{equation}
where $O_{i}$ is the observed number of data points in $i$th bin and $E_{i} = n p_{i}$ is the expected number of data points falling into $i$th bin, here $p_{i}$  is the probability that observation falls into $i$th bin. To apply the test to PRNGs [0,1] interval is divided into $k$ bins and the $\chi^{2}$ statistic is computed noting that $p_{i} = 1/k$. If $H_{0}$ is true then statistics defined in (\ref{chisquareStat}) computed for random samples follows chi-squared distribution with $\nu = (k - 1)$ degrees of freedom

\begin{equation}\label{chisquareDist}
g_{\nu}(y) = \frac{2^{-\frac{\nu}{2}} e^{-\frac{y}{2}} y^{\frac{\nu}{2}-1}}{\Gamma(\frac{\nu}{2})},
\end{equation}
where $\Gamma(\nu)$ is gamma function. It is useful to introduce new random variable $x = \frac{y}{\nu}$ and consider the PDF of $x$ denoted as $f_{\nu}(x)$. This enables to get rid of small numbers in PDFs when $\nu$ is big. Since $1 = \int{f_{\nu}(x) dx} = \int{g_{\nu}(y)} dy$ it follows that 

\begin{equation}\label{RedchisquareDist}
f_{\nu}(x) = \nu g_{\nu}(\nu x) =  \frac{2^{-\frac{\nu}{2}} \nu^{\frac{\nu}{2}} e^{-\frac{x}{2 \nu}} x^{\frac{\nu}{2}-1}}{\Gamma(\frac{\nu}{2})}
\end{equation}
The new random variable introduced in (\ref{RedchisquareDist}) is called reduced chi-square. The  distribution of reduced chi-square for MIXMAX and Mersenne Twister is shown in Fig.~\ref{fig6} in comparison with theoretical expectation.

\section{\textbf{Serial tests.}} 

The serial test also known as chi-square test of independence is multidimensional analogue of chi-square test which checks the independence between two or more random variables \cite{Knuth, Tezuka}. When serial test is applied to PRNGs one divides random sequence into groups of non-overlapping $d$-tuples ($x_{id}, x_{id + }, ... , x_{id + k - 1}$), where $i = 1, 2, ..., \frac{n}{d}$, hence the elements of $d$-tuple are considered as realizations of $d$ random variables and  the relationship between them is of interest . If $x_{i}$s are $U[0,1]$ random variables then $k$-tuples are uniformly distributed in $[0,1]^d$. To check this each dimension of unit hypercube is divided into $k$ bins and the data of $d$-tuples is binned into $[0,1]^d$. Now chi-square statistic (\ref{chisquareStat}) is applied to this data comparing the number of observations falling in each sub-hypercube with theoretical expectation: $E_{i,j,...,d} = np_{i,j,...,d}$, where the joint probability $p_{i,j, ...,d}$ of $d$-dimensional data point to fall into ($i,j,...,d$) sub-hypercube is the product of probabilities of each individual coordinate to fall into appropriate bin, which is the condition of independence.

\begin{equation}
p_{i,j, ...,d} = \prod_{n=1}^{d} p_n =  \left(\frac{1}{k}\right)^d
\end{equation}
Unlike the non-overlapping tuples, the overlapping $d$-tuples of random sequence fall on neighboring parallel planes. The largest distance between adjacent parallel hyperplanes is called the spectral test statistic \cite{Knuth, lecuer1, lecuer2, Dieter, Fishman, Hellekalek}. If the largest distance is small then it implies that overlapping $d$-tuples are more uniformly distributed in unit hypercube, therefore PRNG is considered good.

\begin{figure*}[htbp]
\centering
    \includegraphics[width=10cm]{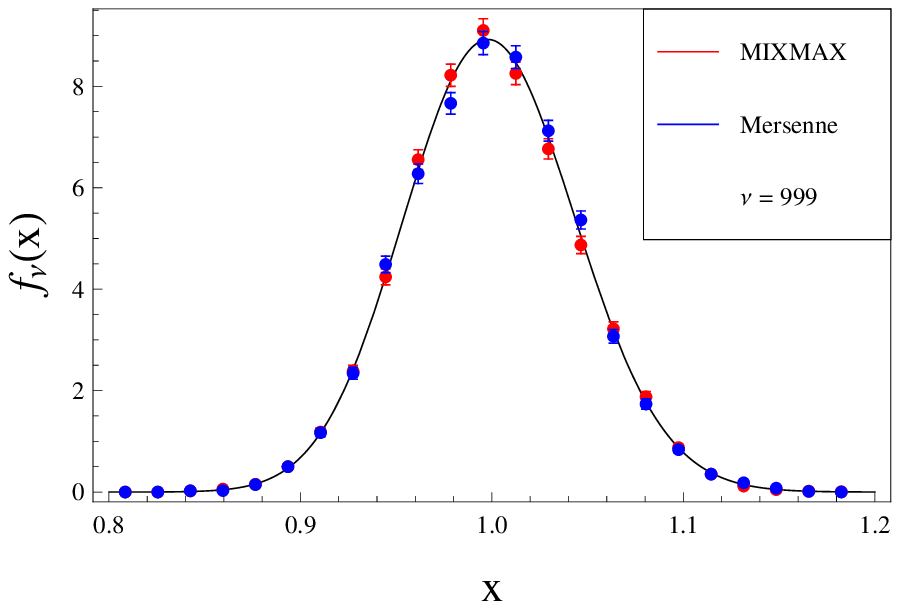}
    \includegraphics[width=10cm]{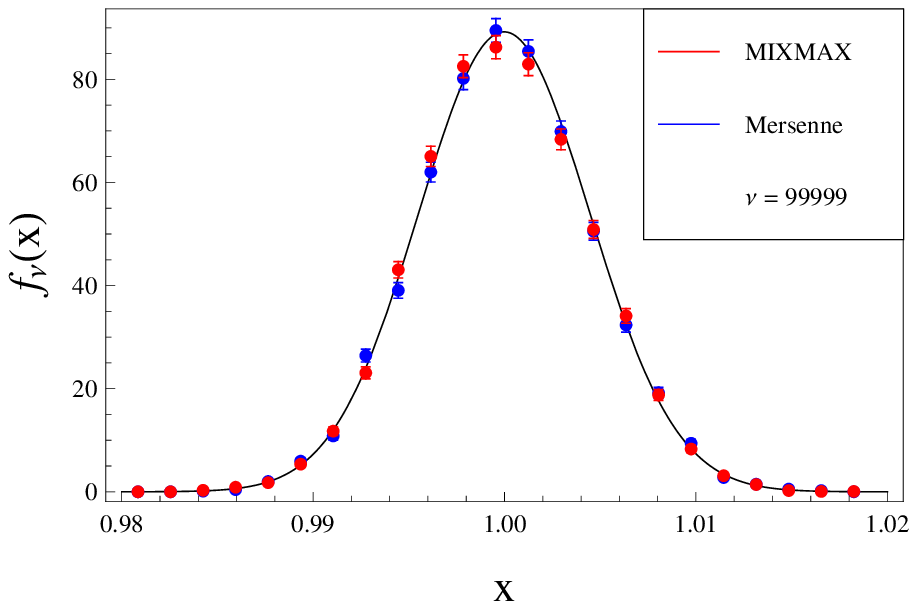}
    \caption{Comparing denisty histogram of reduced chi-square with the test distribution (\ref{RedchisquareDist}) for 3-dimensional(up) and 5-dimensional(down) cases.}
    \label{fig7}
\end{figure*}

The Fig.~\ref{fig7} shows $d = 3$ and $d = 5$ dimensional cases of serial test, where each dimension is divided into $k = 10$ bins and the histogram of reduced chi-square test statistics is compared with the  distribution of (\ref{RedchisquareDist}) with appropriate degrees of freedom.  The reduced chi-square distribution reveals no significant distinction between MIXMAX and Mersenne Twister.
Note that the serial test here is applied to single random stream and measures correlations between adjacent random tuples. This test can be also applied to different streams to check the independence between them.
 
\section{\textbf{Parallel streams of MIXMAX.}} 

All tests described in previous sections use one stream generated by PRNG. However, in multiprocessor stochastic computations it is important to have uniformly distributed and statistically independent simultaneous random streams partitioned across the processors \cite{parallelApplication1, parallelApplication2, parallelApplication3, parallelApplication4, parallelTesting_parallelApplication5_parallelizationTechnique3NoSkip}. Different parallelisation approaches of PRNGs have been studied in literature \cite{parallelizationTechnique1_Skip,parallelizationTechnique2_TwoApproach,parallelizationTechnique3,parallelTesting_parallelApplication5_parallelizationTechnique3NoSkip}. One trivial technique for parallelisation is to take random seeds on each processor, but since every PRNG has finite number of states, one should be careful in order to avoid possible overlapping between different streams. MIXMAX has very large state space, therefore even taking random seeds on each processor does not affect the independence between multiple streams. Another approach is to take single sequence and partition it into different processors. MIXMAX provides skipping-ahead algorithm which enables to skip forward by large amount of numbers in sequence, this technique guarantees the non-collision of partitioned streams \cite{parallelizationTechnique1_Skip}. 
The check for randomness of each individual stream can be done via the standard chi-square or K-S tests. The test of independence of multiple streams can be done via  parallel version of serial test simply forming $d$-tuples of random numbers taken from each of $d$ streams at a time. However, it is not practical to test empirically all random streams when the period is very large, different techniques for testing parallel streams can be found in  \cite{parallelTesting_parallelApplication5_parallelizationTechnique3NoSkip}. 

The serial test up to dimensions $d = 7$ has been performed and it is observed that multiple streams of MIXMAX are statistically independent which is also guaranteed by underlying theory of MIXMAX.

One can analyze the independence and uniformity of parallel streams using the fact that the sum of $n$ independent $U[0,1]$ random variables follow Irwin–Hall distribution of order $n$ \cite{IrwinHall}.

Under $H_{0}$ if each stream of PRNG is generated from uniform distribution then the random sequence resulted from element-wise addition of multiple streams has Irwin–Hall distribution. 
The Irwin–Hall distribution has the following form

\begin{equation}
\label{IrwinHallDist}
f_{n}(x) = \frac{1}{2(n-1)!} \sum_{i=1}^{n} (-1)^{i} {{n}\choose{i}}  (x - i)^{n - 1} sgn(x - i),
\end{equation}
where $sgn(x - i)$ is $sign$ function. When $n = 2$, then (\ref{IrwinHallDist}) reduces to the well known triangular distribution

\begin{equation}
f_{2}(x) =
\begin{cases}
  x, & 0 \leq x < 1  \\
  2-x, & 1 \leq x \leq 2
\end{cases}
\end{equation}

\begin{figure}[H]
\centering
\includegraphics[width=10cm]{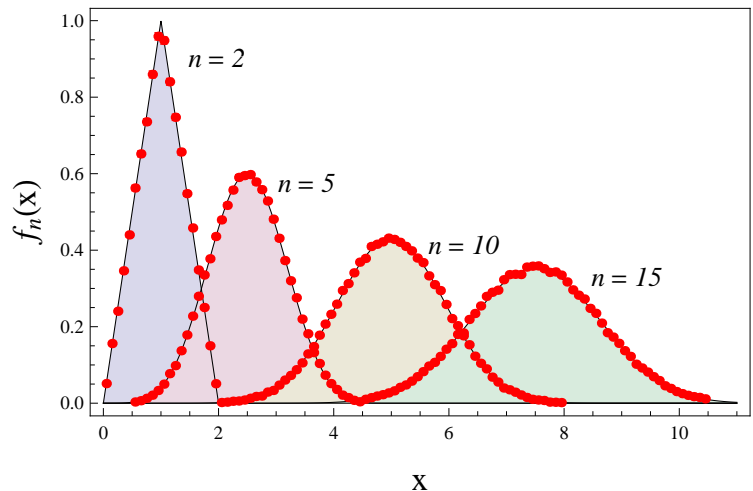} 
\caption {The distribution of Irwin-Hall of order (2,5,10,15) compared with density histogram of data.}
\label{IrwinHallFigure}
\end{figure}

In  Fig.~\ref{IrwinHallFigure} visual comparison of the data with Irwin–Hall distribution is shown, where up to 15 random streams is taken to form the sum. 

To check visual consistency of the histogram and  model prediction in Fig.(\ref{IrwinHallFigure}) chi square test is applied for comparison with Irwin-Hall distribution. The Table \ref{tab:2} represents {\it p}-values of chi square statistic.

\begin{table}[H]
\begin{center}
\vspace{0.3cm}
\begin{tabular}{|c"c |c| c| c|}
\hline
$\textcolor{white}{}$ ~~~Test~~~ &~~n = 2~~&~~n = 5~~&~~n = 10~~&~~n = 15~~\\
\hline
$\textcolor{black}{\textbf{{\it p}-value}}$ & $\textcolor{black}{0.76}$ & $\textcolor{black}{0.97}$ & $\textcolor{black}{0.39}$ &
$\textcolor{black}{0.37}$\\
\hline
\end{tabular}
\caption{{\it p}-values from chi-square test.} 
\label{tab:2}
\end{center}
\end{table}

\section{Summary}

This paper presents the study of newly released MIXMAX PRNG. The various statistical tests including very high quality TestU01 have been used to check quality of MIXMAX compared with other generators, mainly with Mersenne Twister. The results show that MIXMAX is not inferior to Mersenne Twister and even better in sense of speed and period.

\section*{\textbf{ACKNOWLEDGEMENTS}} The authors would like to thank George Savvidy for very useful discussions and comments. This project has received funding from the
European Union's Horizon 2020 research and innovation programme under the Marie Sk\'lodowska-Curie grant agreement No 644121.

\end{document}